\documentclass{bmcart}

\RequirePackage{hyperref}
\usepackage[utf8]{inputenc} 

\def\includegraphics{}

\usepackage{graphicx,graphics,epsfig,natbib}
\usepackage{amsmath}

\startlocaldefs
\endlocaldefs

\begin{document}

\begin{frontmatter}

\begin{fmbox}
\dochead{Research}


\title{A statistical normalization method and differential expression analysis for RNA-seq data between different species}


\author[
   addressref={aff1},
]{\inits{YZ}\fnm{Yan} \snm{Zhou}}
\author[
   addressref={aff1},
]{\inits{JDZ}\fnm{Jiadi} \snm{Zhu}}
\author[
   addressref={aff2},
]{\inits{TJT}\fnm{Tiejun} \snm{Tong}}
\author[
   addressref={aff3},
]{\inits{JHW}\fnm{Junhui} \snm{Wang}}
\author[
   addressref={aff1},
]{\inits{BQL}\fnm{Bingqing} \snm{Lin}}
\author[
   addressref={aff1},
   corref={aff1},
email={zhangjunstat@gmail.com}
]{\inits{JZ}\fnm{Jun} \snm{Zhang}}

\address[id=aff1]{
  \orgname{College of Mathematics and Statistics, Institute of Statistical Sciences, Shenzhen University}, 
  \city{Shenzhen},                              
  \postcode{518060},                                
  \cny{China}                                    
}
\address[id=aff2]{
  \orgname{Department of Mathematics, Hong Kong Baptist University}, 
  \city{Hong Kong}                              
}
\address[id=aff3]{
  \orgname{Department of Mathematics, City University of Hong Kong}, 
  \city{Hong Kong}                              
}



\end{fmbox}


\begin{abstractbox}

\begin{abstract} 
\parttitle{Background} 
High-throughput techniques bring novel tools but also statistical challenges to genomic research.
Identifying genes with differential expression  between different species is an effective way to discover evolutionarily conserved
transcriptional responses.
To remove systematic variation between different species for a fair comparison, the normalization procedure serves as  a  crucial  pre-processing  step that adjusts for the varying sample sequencing depths and other confounding technical effects.
\parttitle{Results} 
In this paper, we propose a scale based normalization (SCBN) method by taking into account the available knowledge of conserved orthologous genes and hypothesis testing  framework. Considering the different gene lengths and unmapped genes between different species, we formulate the problem from the  perspective of hypothesis testing and search  for the optimal scaling factor that minimizes the deviation between the empirical and   nominal type I errors.
\parttitle{Conclusions}
Simulation studies show that the proposed method performs significantly better than the existing competitor  in a wide range of
settings.
An RNA-seq dataset of different species  is also  analyzed and it coincides with the conclusion  that  the proposed method outperforms  the
existing method. For practical applications, we  have also developed an R package
named  ``SCBN" and are available at http://www.bioconductor.org/packages/devel/bioc/html/SCBN.html.

\end{abstract}



\end{abstractbox}
%

\end{frontmatter}



\section*{Background}
High-throughput techniques provide  a high revolutionary technology to replace
hybridization-based microarrays for gene expression  analysis \cite{Mardis2008, Wang2009, Morozova2009}.
The next-generation  sequencing has evoked a wide range of applications, e.g., splicing variants \cite{wan:san:08, sul:sch:ric:08} and single nucleotide polymorphisms \cite{wan:sun:08}. In particular, RNA-seq has become an attractive alternative to detect genes with  differential  expression (DE) between different species, which is used to explore the evolution of gene expression levels in mammalian organs \cite{brawand:11} and the effect of gene expression levels in medicine. As one example, gene expression analyses performed in model
species such as mouse is commonly used to study human diseases \cite{Ala2009}, including  cancer \cite{Segal2005, Sweet2005} and hypertension \cite{Marques2010}.

 For different species, several studies have emerged  in the recent literature to compare the gene expression levels in different organisms using microarrays or RNA-seq data.
 Liu et al.\cite{Liu2011} reported a systematic comparison of RNA-seq for detecting differential gene expression between closely related species. Lu et al.\cite{Lu2010} developed some probabilistic graphical models  and use them to analyze the gene expression between different species.
  Kristiansson et al.\cite{Kristiansson2013} proposed  a new statistical method for meta-analysis of gene expression profiles from different species with RNA-seq data. For different species, the RNA-seq experiments will result in not only different gene numbers and gene lengths, but also different read counts, i.e., sequencing depths.  To  make the expression levels of orthologous genes comparable between different species, normalization is a crucial step  in the data processing procedure.

\begin{figure*}[!hpt]
\includegraphics[width=4.8in, height=2in]{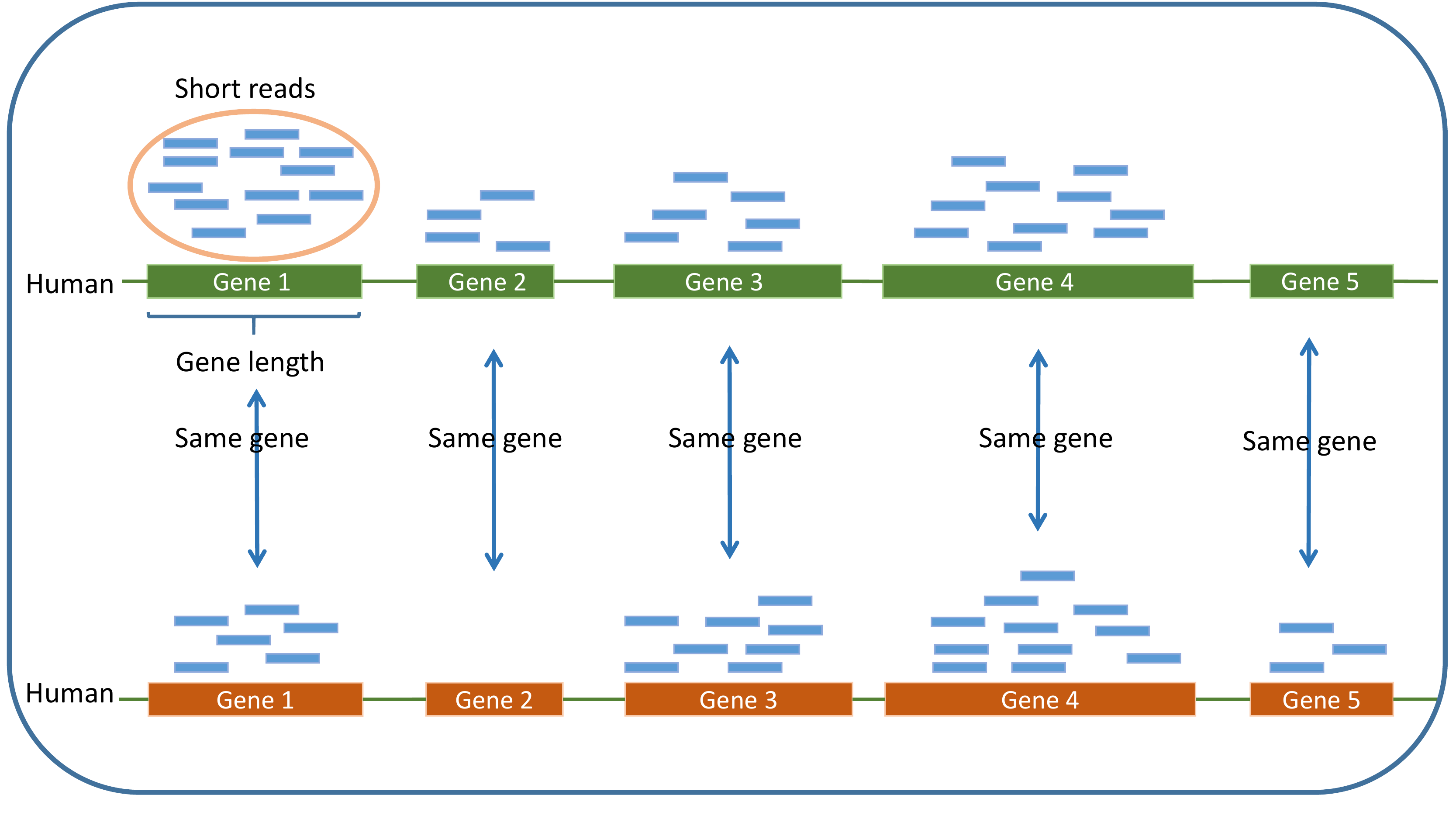}
\includegraphics[width=4.8in, height=2in]{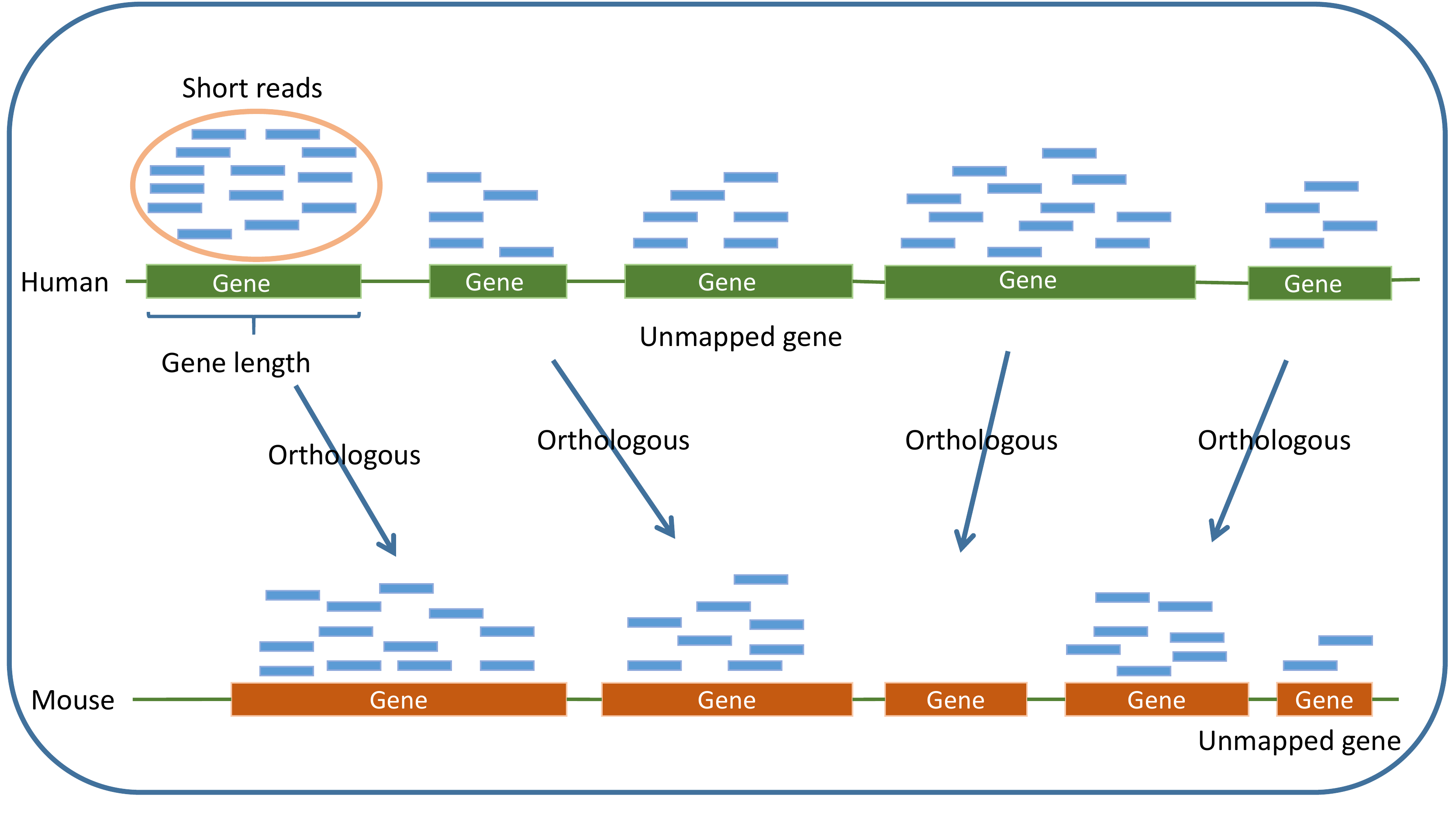}
\caption[Illustration]{The first panel shows the same genes of different human samples, and the second panel shows the orthologous genes in human and mouse.  }\label{fig:samesp}
\end{figure*}

The main purposes of normalization are  to remove the systematic variation and to reduce  the noise in the data. In the case of one species (see the first panel of Fig~\ref{fig:samesp}), various  normalization methods have been developed in the last decade, e.g.,  modifying the gene expression  with a global factor \cite{Bol:Iri:03, Mar:Mas:08, Rob:Smy:06, Bul:Pur:09}.
Mortazavi et al.\cite{Mor:Wil:08} transformed RNA-seq data to reads per kilobase per million mapped (RPKM). Robinson et al.\cite{rob:osh:10, robinson2010edger} proposed a weighted trimmed mean of log-ratios method (TMM). Zhou et al.\cite{Zhou2017a} developed a hypothesis testing based  normalization  (HTN) method by utilizing  the available knowledge of housekeeping genes, and   showed that the HTN  method  is more robust than   TMM  for analyzing RNA-seq data.

We note, however, that normalization  of RNA-seq data with different species is more difficult  than that with  same species. For different species,  we need to consider not only the total read counts but also the different gene numbers and gene lengths (see the second panel of Fig~\ref{fig:samesp}).
 To the best of our knowledge,  few studies are  available in the literature for normalizing RNA-seq data with different species. As a routine method for  normalization, one often standardizes the data with  different species by scaling their total number of reads to a common value.
For instance, Brawand et al.\cite{brawand:11} used RPKM  in \cite{Mor:Wil:08} to normalize   RNA-seq data with different species. Specifically, they  first identified
the most conserved 1000 genes between species and then assessed their median expression levels in each species among the genes with
expression values in the interquartile range for different species. Lastly, they  derived the scaling factors that adjust those median values to a
common value.

 In this paper, we extend the HTN method to the setting of different species. As described in \cite{Zhou2017a},  HTN is a normalization method  under different sequence depths for the same species, and its performance outperforms other normalization methods. Based on hypothesis testing framework, it transforms the problem to find the scaling factor in normalization. By utilizing the available knowledge of housekeeping genes, it get the optimal scaling factor by minimizing the deviation between the empirical and the nominal type I error. However, the HTN method cannot be directly applied to normalize the different species dataset, because of the assumption of  the same number and length.
By extending the HTN  method  to the setting of different species, we develop a scale based normalization (SCBN) method by utilizing the available knowledge of conserved orthologous genes and the hypothesis testing framework. Here, we define conserved orthologous genes for different species instead of housekeeping genes. It is noted  that the  normalization scaling factor  is stable for different confidence levels in the hypothesis test in both simulation studies and real data analysis.

The rest of the paper is organized as follows. we first
introduce the new SCBN method in detail in Section ``Materials and Methods". In Section ``Simulation Studies", we conduct simulation studies to assess  the performance of the SCBN method and  also compare it with  the  existing method.  In Section ``Real Data Analysis", we apply the SCBN method   to a real dataset with human and mouse  to  demonstrate  its superiority over the existing method. The  paper is concluded in Section ``Discussion" with some discussions and future work.

\section*{Materials and Methods}
In the following section,   we  propose a novel normalization method  for  RNA-seq data with  different species  by utilizing  the available knowledge of conserved orthologous genes and the hypothesis testing framework.

\subsection*{\bf Notations and model}
Let $G=\{g_1,g_2,\ldots,g_n\}$ be  the complete set of   genes from two different species, and $G_0$ be the set of  one-to-one orthologous genes that are  to be tested for differential expression.  For  species $t=1$ or $2$, let $X_{g_{k}t}$  be the random variable that represents the count of reads mapped to the orthologous gene $g_{k} \in G_0$,  and  $x_{g_{k}t}$ be the observed value  of $X_{g_k t}$. Accordingly, the   total number of orthologous reads for species $t$ is  $N_t=\sum_{g_{k} \in G_0}x_{g_{k}t}$. For ease of presentation,  our normalization method is presented   for  the setting of  one sample in  each species  only. Our proposed  method, however,  can  be readily  extended to more general settings including multiple samples for  each  species. For gene $g_k$ in species $t$,  we  consider the   mean model:
\begin{equation}
		E(X_{g_{k}t}) = \frac{\mu_{g_{k}t}L_{g_{k}t}}{S_t}N_t,\label{eq:1}
\end{equation}
where $\mu_{g_{k}t}$ are the true expression levels,  $L_{g_{k}t}$  are the true  gene lengths,  and $S_t=\sum_{g_{k} \in G_0}\mu_{g_{k}t}L_{g_{k}t}$ is the total expression output of all orthologous genes in species $t$.
Note that, since $L_{g_{k}t}$ are often  different between species, we have included  them  in  model (\ref{eq:1}) to alleviate the bias in gene lengths.

\subsection*{\bf Novel normalization method}
We  propose a novel normalization method by employing the available knowledge of conserved orthologous genes and the hypothesis testing  framework. Specifically, we  choose a scale to minimize the deviation between the empirical and  nominal type I errors in RNA-seq data based on the hypothesis test.

To detect differential expressions of orthologous genes between two species, for each $g_k\in G_0$,  we  consider the hypothesis
\begin{eqnarray*}
H_{0}^{g_k}: \mu_{g_{k}1}=\mu_{g_{k}2}\quad {\rm versus} \quad H_{1}^{g_k}: \mu_{g_{k}1}\neq \mu_{g_{k}2}.
\end{eqnarray*}

We further  assume  that the reads mapped to the orthologous genes are  Poisson random variables
   with $\lambda_{g_{k}1}=E(X_{g_{k}1})$ and $\lambda_{g_{k}2}=E(X_{g_{k}2})$. Then under  model (\ref{eq:1}), the  hypothesis is equivalent  to
\begin{align}
H_{0}^{g_k}:\lambda_{g_{k}1}=\frac{L_{g_{k}1}}{L_{g_{k}2}}\frac{N_1}{N_2}c\lambda_{g_{k}2}\quad
{\rm versus}\ \  H_{1}^{g_k}:\lambda_{g_{k}1}\neq\frac{L_{g_{k}1}}{L_{g_{k}2}}\frac{N_1}{N_2}c\lambda_{g_{k}2},
\label{eq:2} 
\end{align}
where  $c=S_2/S_1$  is   the scaling factor for normalization.

Given that  $X_{g_{k}1}+X_{g_{k}2}=n_{g_{k}}$  with  $n_{g_{k}}$   a fixed integer, the random variable $X_{g_{k}1}$ follows a binomial distribution with the conditional probability density function as
   \begin{eqnarray*}
	&& P(X_{g_{k}1}= x_{g_k 1}\big|{X_{g_{k}1}+X_{g_{k}2}}=n_{g_{k}})
	=\frac{n_{g_{k}}!}{x_{g_{k}1}!(n_{g_{k}}-x_{g_{k}1})!} (p_0^{g_{k}})^{x_{g_{k}1}}
	 (1-p_0^{g_{k}})^{n_{g_{k}}-x_{g_{k}1}},
	 \end{eqnarray*}	
where	
$$p_0^{g_{k}}=\frac{\lambda_{g_{k}1}}{\lambda_{g_{k}1}+\lambda_{g_{k}2}}=
   \frac{cL_{g_{k}1}N_1}{L_{g_{k}2}N_2+cL_{g_{k}1}N_1}$$
is the probability of success  under the null hypothesis of (\ref{eq:2}). For the above model, the $p$-value of the test is
\begin{align}
p_{g_{k}}(c)&=P( | X_{g_{k}1}-n_{g_{k}} p_0^{g_{k}}|\geq | x_{g_{k}1}-n_{g_{k}} p_0^{g_{k}}|\big|n_{g_{k}})\notag  \nonumber\\
&=P(|(1+\frac{L_{g_{k}1}}{L_{g_{k}2}}\frac{N_1}{N_2}c)X_{g_{k}1}-\frac{L_{g_{k}1}}{L_{g_{k}2}}\frac{N_1}{N_2}cn_{g_{k}}| \geq  \nonumber\\ &~~~~~|(1+\frac{L_{g_{k}1}}{L_{g_{k}2}}\frac{N_1}{N_2}c)x_{g_{k}1}-\frac{L_{g_{k}1}}{L_{g_{k}2}}\frac{N_1}{N_2}cn_{g_{k}}|\big|n_{g_{k}}).
\label{eq:3}
\end{align}
Note that the $p$-value in (\ref{eq:3}) is a function of the scaling factor $c$ under the condition $X_{g_{k}1}+X_{g_{k}2}=n_{g_{k}}$. To search for the optimal $c$ for  normalization, we apply  the following two  questions as criteria.  (i) Does the normalization method improve the accuracy of DE detection, i.e.,  whether or not it will decrease  the false discovery rate (FDR) of the tests?   (ii) Does the normalization method  result in a lower technical variability or specificity?  For multiple testing, Storey\cite{Stor:JD:03} pointed out  that different hypothesis tests will result in different significant regions. To  transform these tests into a common space, the $p$-value is a natural way to do so  with respect to the positive false discovery rate (pFDR). By taking the number of set $G_0$  identical hypothesis tests, the pFDR is defined as follows:
\begin{align}
\mbox{pFDR}_{g_{k}}&=\!\frac{P(H_0;c)P( R_{g_k}\mid H_0; c)}{P( R_{g_k};c)} \nonumber\\
&\!=\!\frac{P(H_0;c)P( R_{g_k}\mid H_0;c)}{\!P(H_0;c)P( R_{g_k}\mid
H_0;c)+P(H_1;c)P( R_{g_k}\mid
H_1;c)\!},
\label{eq:31}
\end{align}
where $\alpha$ is the significance level and $R_{g_k}=\{p_{g_k}(c) < \alpha \}$ is the rejection region. By (\ref{eq:31}), the pFDR of gene $g_k$ is a function of both $\alpha$  and $c$. Given the values of $\alpha$ and $c$, we can apply the empirical distributions to
estimate $P(R_{g_k}|H_0;c)$ and $P(R_{g_k}|H_1;c)$. Let $V_0$ and $V_1$ be the sets of non-DE genes and DE genes in $G_0$, respectively.    Then, $\mbox{pFDR}_{g_{k}}(\alpha;c)$ can be  estimated as

\begin{eqnarray*}
  \widehat{\mbox{pFDR}}_{g_{k}}=\!\frac{P(H_0;c){\widehat P}( R_{g_k}\mid H_0;c)}{P(H_0;c){\widehat P}( R_{g_k}\mid H_0;c)+P(H_1;c){\widehat P}( R_{g_k}\mid
H_1;c)\!},
\end{eqnarray*}
where
$${\widehat P}(R_{g_k} \mid H_0;c)=\frac{1}{n_0}\sum_{g_k\in V_0}I(p_{g_{k}}(c)<\alpha|H_0;c)$$ for any ${g_k\in V_0}$, and
$${\widehat P}(R_{g_k}\mid H_1;c)=\frac{1}{n_1}\sum_{g_k\in V_1}I(p_{g_{k}}(c) <\alpha|H_1;c)$$ for any ${g_k\in V_1}$,
where $I(\cdot)$ is the  indicator function, and  $n_0$ and $n_1$ represent the cardinalities of $V_0$ and $V_1$, respectively.

When all non-DE genes in $V_0$ are given, we  can perform our new  normalization by determining the  optimal scaling factor that minimizes the value of pFDR. For real data, however, it is not uncommon  that only  a small proportion of non-DE genes are known a priori by  background knowledge. In this paper, we assume that a set of conserved orthologous  genes  between species are given in advance, which may either   be reported in other studies or be selected by a certain biological measure \cite{brawand:11, chen:pai:14}.
For the given set $H$ of  conserved orthologous genes that  are considered as non-DE genes for its stability between species, we search  for the optimal scaling factor by minimizing  the deviation between the empirical and  nominal type I errors. Let $m$ be the   number of genes in  the set $H$.
Given the true value of $c$, the $p$-values of the tests for the conserved orthologous genes  follow a uniform distribution on interval $(0,1)$. That is, for the specified $\alpha$ and $c$, the value of   $\sum_{g_k \in H}(1/m)I(p_{g_{k}}(c)<\alpha|H_0;c)$ should be around  the nominal level at $\alpha$. In our method, we define the   optimal scaling factor as $c_{\rm opt}$ that  minimizes the objective function $\mid\sum_{g_k \in H}(1/m)I(p_{g_{k}}(c)<\alpha|H_0;c)-\alpha\mid $; that is,
\begin{align}
c_{\rm opt}=\underset{c>0}{\mbox{argmin}} \big|\sum_{g_k \in H}\frac{1}{m}I(p_{g_{k}}(c)<\alpha|H_0;c)-\alpha \big|. \label{eq:33}
\end{align}

Finally, to estimate the optimal scaling factor defined in (\ref{eq:33}), we apply  a grid search method and denote the best estimate as $\hat c_{\rm opt}$. For convenience, we refer to the proposed scale based normalization method as  the SCBN method.

\section*{Simulation Studies}
For a fair comparison, we generate the simulation datasets  following the  settings in \cite{rob:osh:10}, but with the structure of different species  rather than  same species. For different species, we  consider different sequencing depths and  lengths of orthologous genes  to  generate  the  datasets,  including the  DE genes, non-DE genes and unmapped genes for two species  to mimic the real scenario.
The unmapped genes represent those genes that exist only in one species. They are different from the unique genes, representing those orthologous genes that exist in both species but are expressed in only one of them.
After setting   the numbers of unique genes  and  unmapped genes,   proportion,  magnitude and   direction of DE genes between two species, we  randomly generate the rate  of a gene expression level to the output of all the orthologous genes from a given empirical distribution of real counts. We set the expected values of the Poisson  distributions from  model (\ref{eq:1}), and then randomly generate simulation datasets  from the respective  distributions.

\begin{figure}[h!]
\centerline{\includegraphics[width=4.8in,height=3.2in]{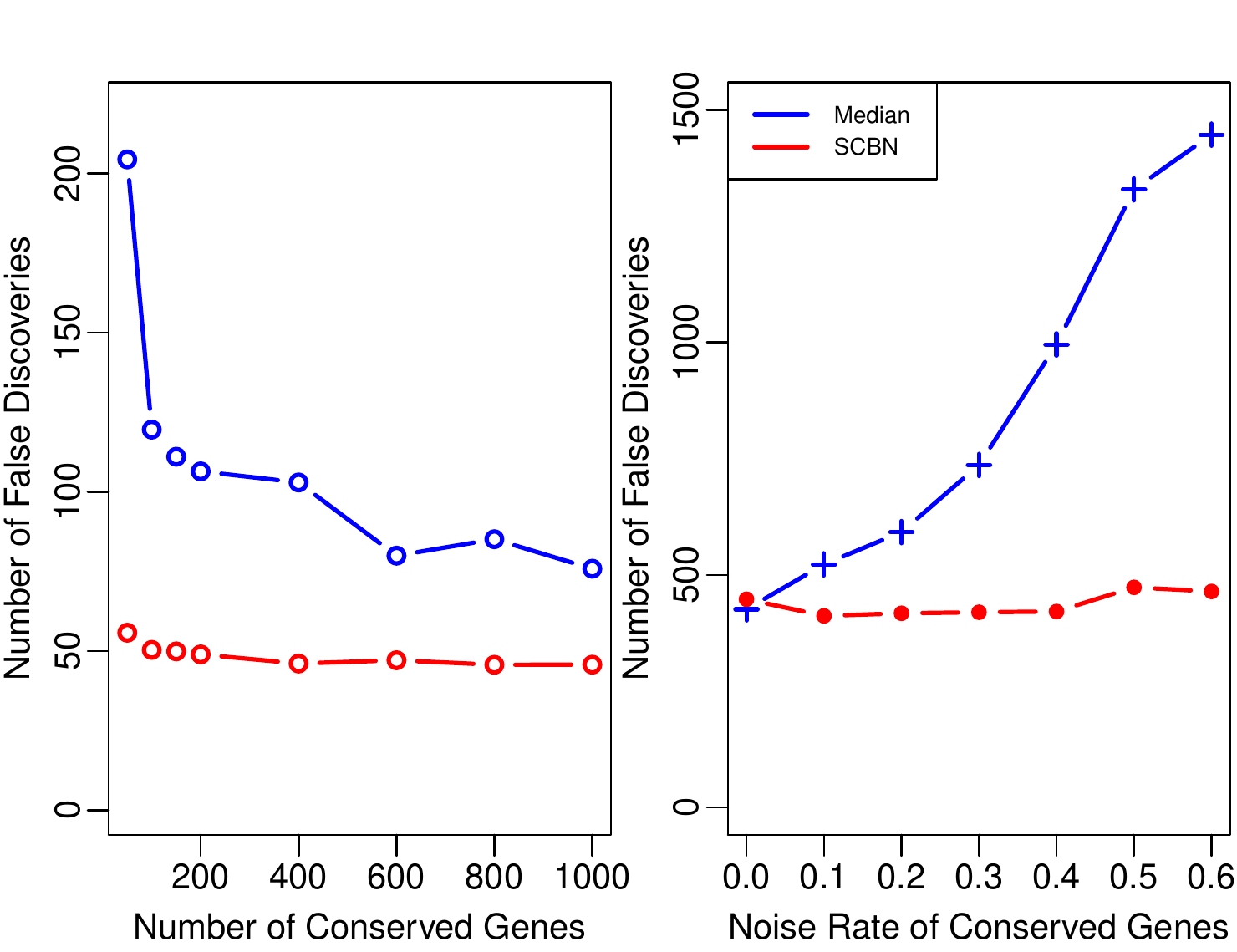}}
\caption[Illustration]{The left panel is the false discovery numbers of the median and SCBN methods with different numbers of conserved genes. The right panel is the false discovery numbers of the two methods with different rates of noise in conserved genes.}\label{fig:1}
\end{figure}

We first evaluate  the stability of the proposed SCBN  method for the fixed parameters. In Study 1, we compare the false discovery numbers of  the SCBN method and the median method with different numbers of conserved genes. We set 10\% of the orthologous genes as DE genes at the  1.2-fold level; of those DE genes, 90\% are up-regulated in the second species, and we set the number of unique genes as 1000 and 2000 for two species, respectively. Besides, we set 2000 and 4000 unmapped genes for two species. With the fixed parameters, we consider the cases where the number of conserved orthologous genes varies from 50 to 1000. In Study 2, the parameters are the same as those in Study 1 except that the fold level of DE genes is increased to 1.5, and we select 1000 conserved genes in each experiment. Then, we investigate the stability of the proposed method when the rates of  noise in conserved genes increase from 0 to 0.6 with step size 0.1. In Study 3, we consider the adjusted M versus A plots in \cite{rob:osh:10}  to compare the scaling factors of two normalization methods when the rates of  noise in conserved genes equal to 0 and 0.4. In this paper, the rate of  noise  means the proportion of DE genes in all the conserved genes. To make it more obvious, we adjust the parameters with  20\% DE genes at the  8-fold level, and 70\% are up-regulated in the second species. The unique genes and unmapped genes are the same as before. In Study 4, we test the stability of the SCBN method by choosing different $p$-values as cutoff. In this study, we consider the cutoff values varying from 0.0001 to 0.6. The parameters are the same as those in Study 1 except that  40\% of genes are differentially expressed.

 Next, we investigate the performance of the SCBN method with several criteria, including the  false discovery number, precision, sensitivity and $F$-score, which were also adopted in  \cite{Lin2014}. In Studies 5 and 6,  the   parameters are kept the same as those in  Study 2. In Study 5, the false discovery numbers of the two normalization methods are shown with different rates of noise in conserved genes, ranging from 0 to 0.5. In Study 6, we compare the   precision, sensitivity  and $F$-score  for the two methods. The precision denotes the rate of true positives in all the predicted positives, the sensitivity represents the rate of true positives in all real positives, and the $F$-score is a metric to overview both  the precision and sensitivity.  Here, we take 0.01  as the $p$-value cutoff.

In Study 7, we compare the performance of the two methods for different rates of DE genes in all orthologous genes. We set the fold change of DE genes as 1.5, the rate of noise in conserved genes as 0.2, and  the rate of DE genes varying  from 0.1 to 0.6. Other parameters are kept the same as those in Study 4.

\begin{figure}[!h]
\centerline{\includegraphics[width=4.8in,height=3.2in]{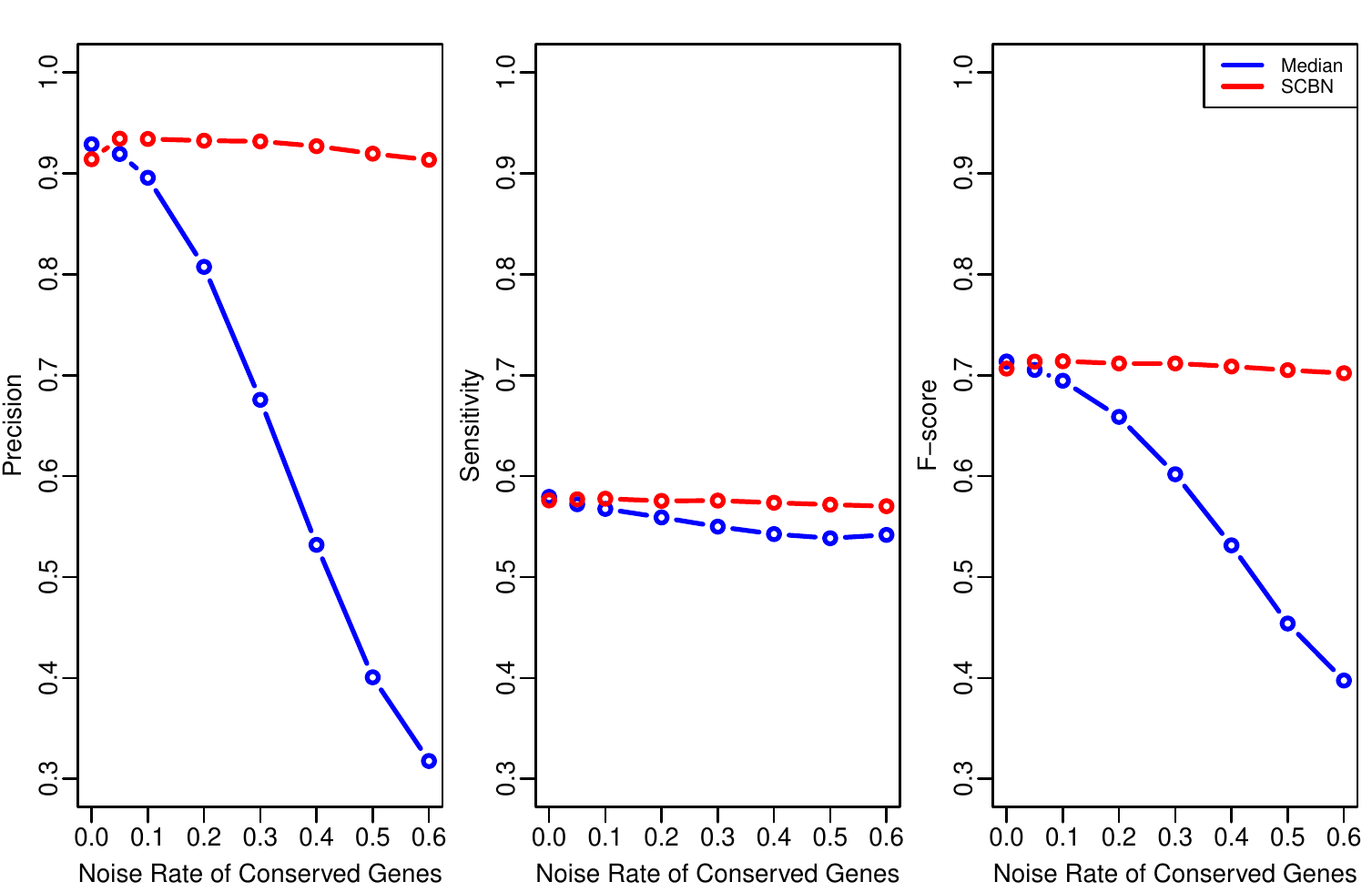}}
\caption[Illustration]{Precision (left), sensitivity (middle) and $F$-score (right)  values of two normalization methods with various rates of noise.}\label{fig:5}
\end{figure}

For each simulated dataset, we compare the false discovery numbers, which are computed by repeating the simulation 100 times, while there are time consuming in each repeat, and averaging over all the repetitions. 
 We report the stability of the SCBN method with various parameters in Fig~\ref{fig:1}. Fig~\ref{fig:5} compares the SCBN method to the median method with false discovery number, precision, sensitivity and $F$-score criteria. The Additional file 1 compares the false discovery numbers with different rates of noise in the selected conserved genes.

  The left panel of Fig~\ref{fig:1} (Study 1) shows that the  false discovery numbers  are reduced as the number of conserved genes increases. Whereas the false discovery numbers of the median method increase drastically when conserved genes become less, the SCBN method is much more robust to the number of conserved genes. Furthermore, the SCBN method performs much better than the median method for each number of conserved genes.
 As shown in the right panel of Fig~\ref{fig:1} (Study 2),  the  false discovery numbers of the SCBN method keep stable, but that of the median method increases gradually as the rate of noise increases. From these two studies, we can see that the SCBN method is   more robust than the median method, especially when the number of conserved gene is small, or the rate of noise  is large.

 In Study 3, the two scaling factors are presented in Additional file 2. From the left panel, the lines of the two normalization methods are close when conserved genes do not include noise. However, as the rate of noise equals to 0.4, the right panel shows the scaling factors for the SCBN method are much closer to the center of non-DE genes. Additional file 3 presents the result of Study 4, which demonstrates the choice of $p$-value cutoffs has no impact on the results of the SCBN method.

 In Study 5, we investigate the overall situations of false discoveries changed with different rates of noise. The results are shown in  Additional file 1, which shows that the two normalization methods have a similar performance
 when all selected conserved genes are non-DE genes. However, the SCBN method outperforms the median method when the rate of noise becomes larger than 0.1. Hence, we conclude that the SCBN method  performs significantly better than the median method when moderate-to-large rates of noise are present.

 \begin{figure}[h!]
\centerline{\includegraphics[width=4.8in,height=4in]{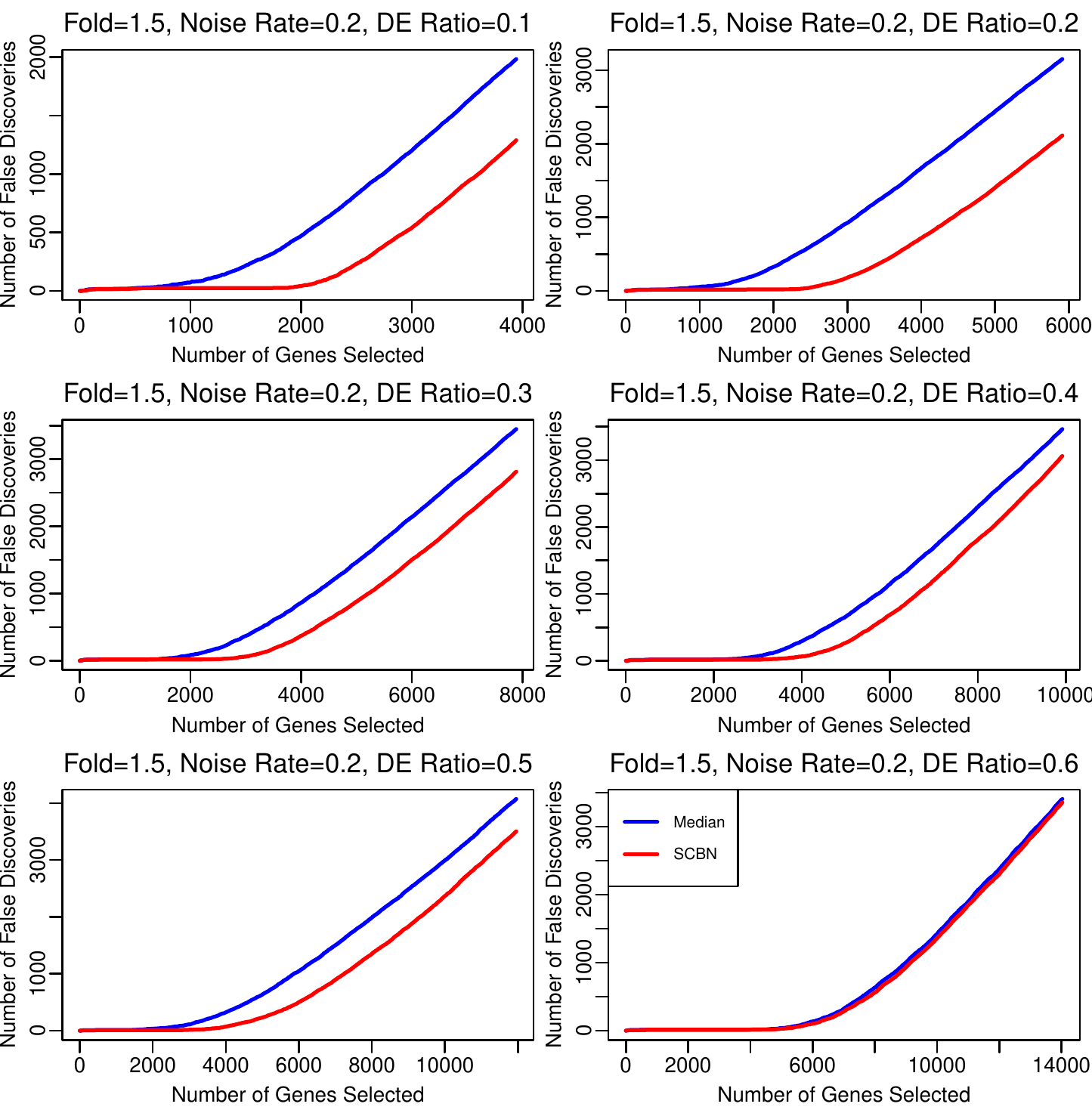}}
\caption[Illustration]{The false discovery number of two normalization methods with DE genes at the rates  of 0.1, 0.2, 0.3, 0.4, 0.5 and 0.6, respectively. }\label{fig:6}
\end{figure}

 Fig~\ref{fig:5} shows the experimental results of precision, sensitivity and $F$-scores. Since $F$-score is the harmonic mean of precision and sensitivity, it is clear that the SCBN method has overall better performance as it achieves higher $F$-scores in most cases. As we can see from the plots, when rate of noise is less than 0.1, the values of sensitivity and $F$-score for two normalization methods are very close. The median method performs slightly better than the SCBN method in precision when conserved genes have no noise or small noise, but its precision decreases enormously with noise increased. For instance, the precisions of the median method are 0.93, 0.68 and 0.32 with conserved genes have 0, 30\% and 60\% of DE genes. The SCBN method has precision values 0.91, 0.93 and 0.91, respectively. It is evident that the median  method depends greatly on the selected conserved genes, including numbers and purity of conserved genes. On contrary, conserved genes have much less impact on the performance of the SCBN method.

 In Study 7, we focus on the impact of the rate of DE genes on two normalization methods. Fig~\ref{fig:6} shows that the SCBN method outperforms the median method for various rates of DE genes, especially when the rate of DE genes is not too large. The result implies that the SCBN method is more sensitive to identify less fold of DE genes  than that of the median method.

\section*{Real Data Analysis}

\begin{figure*}[h!]
\centerline{\includegraphics[width=5in,height=3in]{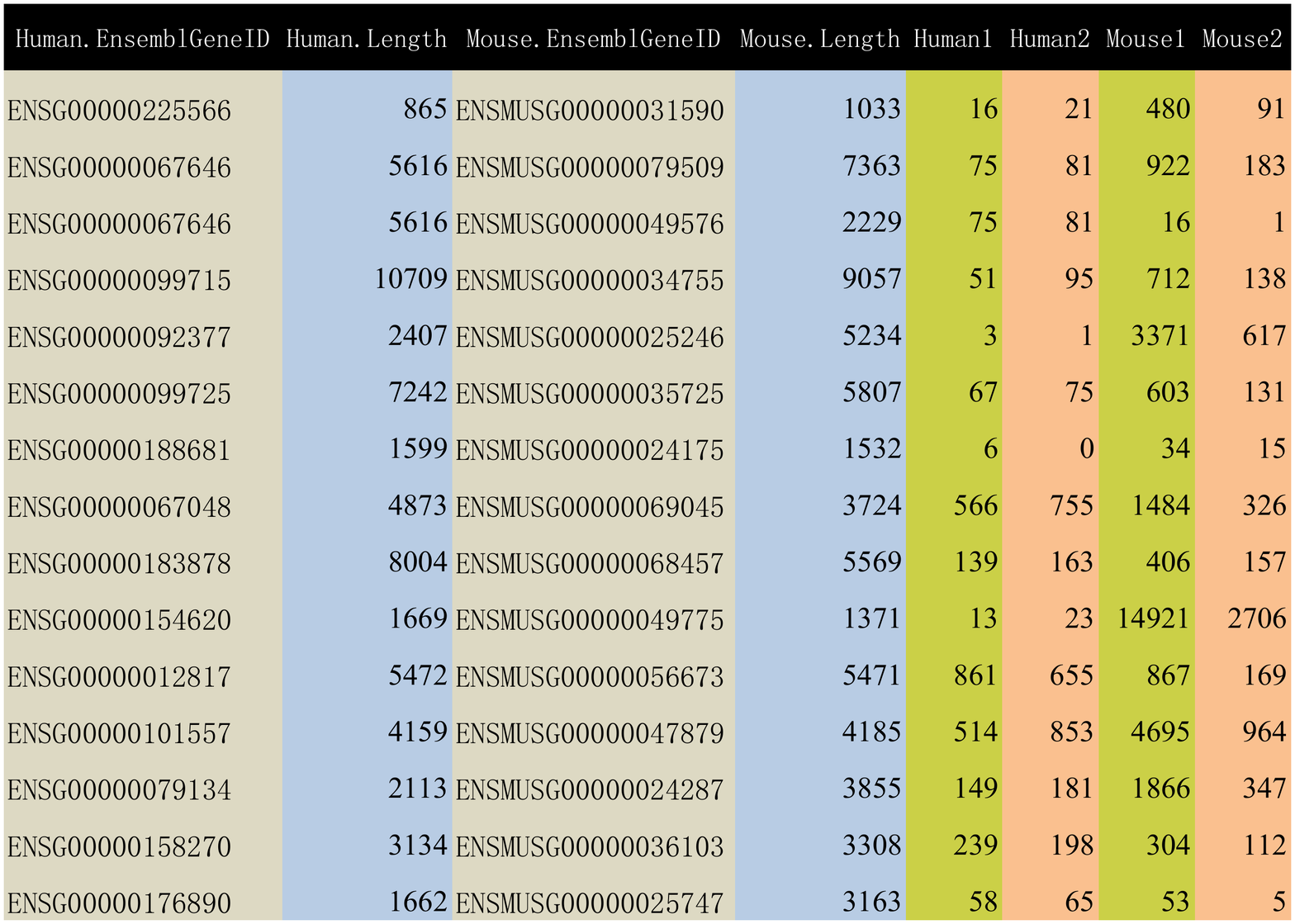}}
\caption{The RNA-seq data of orthologous transcripts in human and mouse.}\label{tab1}
\end{figure*}

We illustrate the usefulness of our  proposed SCBN method in real dataset by the study of  \cite{brawand:11}.  The dataset  consists of two groups of orthologous transcripts in human and mouse, with  respective transcripts lengths and counts of reads (see Additional file 4  for details).
We refer to the human transcripts (GRCh38.p10) and the mouse transcripts (GRCm38.p5) in Ensembl Genes 91 available at http://asia.ensembl.org/biomart/martview/eb8237d1bbc63bbeff285/c0d5a2c3742. There are a total of 63967 transcripts in human and 53946 transcripts in mouse, 27779 of which are orthologous transcripts (see the right panel of Fig~\ref{fig:samesp}).
By excluding the unmatched, duplicated and unexpressed  transcripts, there are  19330 available orthologous  transcripts. Fig~\ref{tab1} shows the expressions of several orthologous transcripts in human and mouse.


As shown in Fig~\ref{fig:samesp}, unlike the case of same species where  the  numbers and lengths of genes are equal to each other,
 different species have different gene numbers and
 thus different gene lengths. Regarding the different lengths of the orthologous transcripts, only 105 transcripts or only 0.54\% of all transcripts, have the same length between human and mouse in  Additional file 5. The average difference of the transcripts lengths  between two species is 1039, and the maximum is 21666 in  Additional file 6.
  The evolutionary process of the eukaryotic genome includes
events such as duplication and recombination, which creates
complicated relationships among genes. As a consequence, the  normalization
methods for same species may not provide a satisfactory
performance or may not even be applicable for different species.
The challenges of
normalization between different species are mainly due to the
different lengths of orthologous genes and the different sequencing
depths due to different platforms.

We get the conserved orthologous genes with a three-step procedure. First, we confirm the orthologous transcripts between human and mouse, by using the BioMart function in the Ensemble to search all human transcripts and filtering  out the genes that do not exist in mouse. Second, according to the orthology quality-controls criterion, we sort the data from the most  conserve  to the least.   Third,   we select  143 most conserved orthologous transcripts  between human and mouse and list them  in Additional file 7.

The most conserved 500 or 1000 orthologous transcripts are likely non-DE transcripts between two species, and we compare the two methods with the first group data. First,  we select  the most 500 or 1000 conserved transcripts with the above steps, and then use the two methods to normalize the sequence data with the 143 conserved transcripts.
 Next, we calculate  $p$-values (see Additional file 8) with adjusted sage.test function. Last, we get   DE transcripts between human and mouse with $p$-value cutoff $10^{-6}$, which are shown  in Table~\ref{table2}. Among the most conserved 500 or 1000 orthologous transcripts,   332 and 647 of them are detected as  DE transcripts by using the SCBN method, in which 48\% and 46\% significantly higher in human, whereas the median method detects 351 and 697 DE transcripts, in which  32\% and 29\% significantly higher in human.   For all orthologous transcripts, the SCBN method detects 9662 DE transcripts, and the median method detects 9910 DE transcripts.
Assuming that the most conserved 500 orthologous transcripts are non-DE transcripts, there are 351  false detected DE transcripts with the median method and 332 false detected DE transcripts with the SCBN method. Then the FDR of the median method is 0.035, which is larger than 0.034 of the SCBN method. For the 1000 conserved transcripts, we get a similar result  that the FDR of the median method (0.070) is also larger than that of the SCBN method (0.067). Therefore, the FDRs of the SCBN method are generally smaller than those  of the median method.

\begin{table}[!ht]
\caption{The number of DE genes between
human and mouse at a cutoff $p$-value $<10^{-6}$ for the
median and SCBN   methods.}
\begin{center}
\begin{tabular}{ccccccccc}
  \hline \hline
                      & Median      &SCBN    & Overlap\\
  \hline
  Higher in human     &4370         &5824      & 2610 \\
  Higher in mouse     &5540         &3838      &2184 \\
  Total               &9910         &9662      & 4794  \\
\hline
 Top conserved genes (500)\\
  Higher in human      &112         &159        &56    \\
  Higher in mouse      &239         &173        &119    \\
  Total                &351         &332        &175    \\
\hline
Top conserved genes (1000) \\
  Higher in human      &201        &300        &87    \\
  Higher in mouse      &496        &347        &240    \\
  Total                &697        &647        &327    \\
\hline

\end{tabular}
\label{table2}
\end{center}
\end{table}

\begin{figure}[h!]
\centerline{\includegraphics[width=3.6in,height=2in]{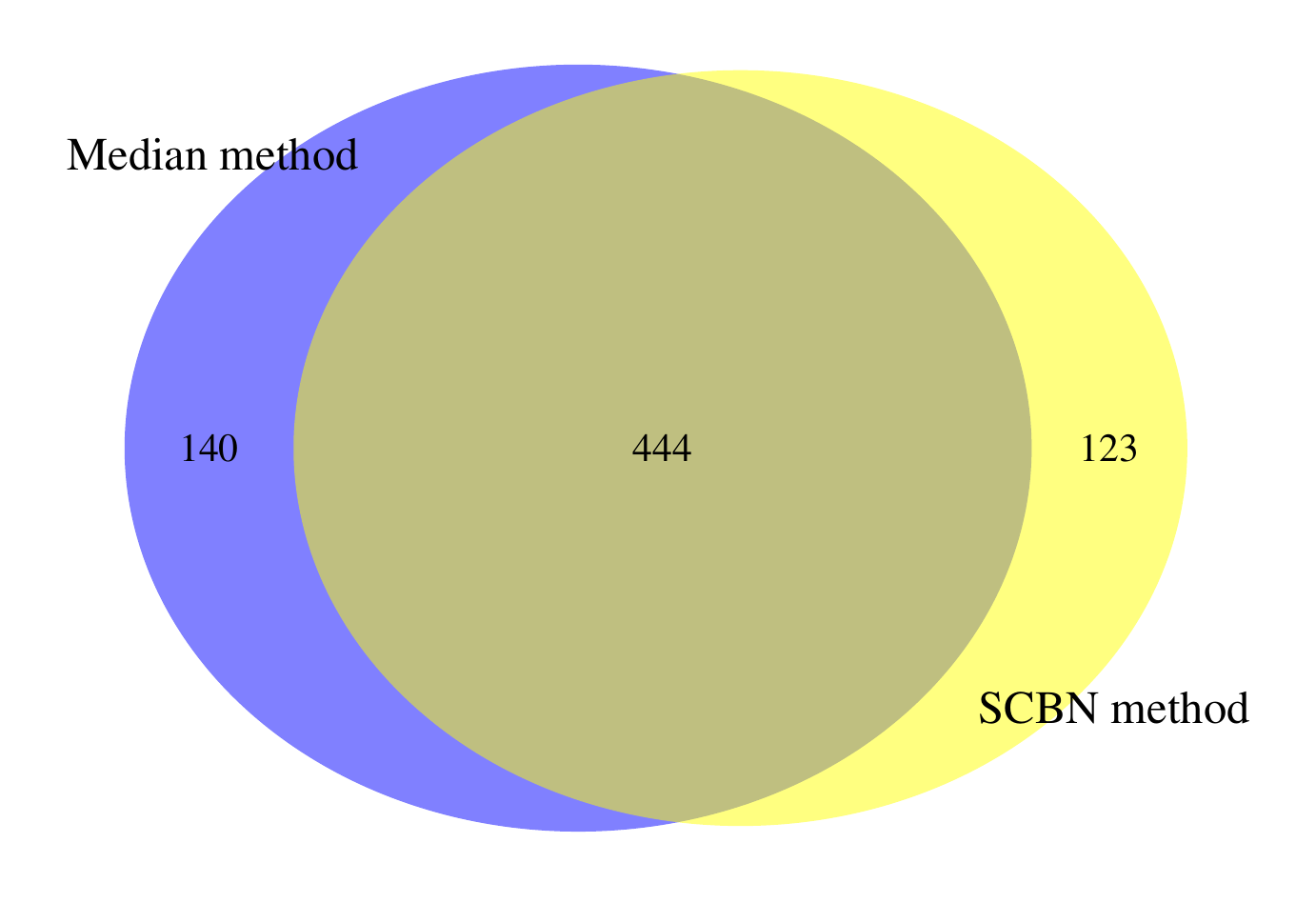}}
\caption[Illustration]{The common genes and the unique DE genes detected by   two normalization methods. }\label{fig:7}
\end{figure}

Next, we compare the accuracy of the two normalization methods by looking deeper into the biological function. We apply the SCBN method to detect the most significant 1000 DE transcripts for each pair comparison between human and mouse,  that is the smallest 1000 $p$-values for each comparison, among which 567 are common. Also, the median method detects 584 common DE transcripts for two species.  Fig~\ref{fig:7} shows the common DE transcripts and the unique DE transcripts of the two normalization methods.
 For the unique transcripts, we refer to NCBI \cite{NCBI:web} to find out which genes are associated with evolution or illness.
 There are 48 of 123 (39.02\%) DE transcripts, which are related to evolution or illness with the SCBN method,  and 43 of 140 (30.71\%) DE  transcripts are related to evolution or illness  with the median method. Specifically, among the unique DE transcripts detected by the SCBN method, we find that `ENSG00000102316' is involved in breast cancer and melanoma, `ENSG00000152137' is involved in the regulation of cell proliferation, apoptosis, and carcinogenesis, and `ENSG00000135744' is associated with the susceptibility to essential hypertension, and can cause renal tubular dysgenesis, a severe disorder of renal tubular development.
 Mutations in  gene `ENSG00000152137' have been associated with different neuromuscular diseases, including the Charcot-Marie-Tooth disease.
We note, however, that   above genes are not included in the 584 most significant DE transcripts detected by the median method. More details are presented in  Additional file 9. The results show that our proposed SCBN method provide a more accurate normalization than the median method in real data analysis.

\section*{Discussion}
Detecting DE genes between different species is an effective way to identify evolutionarily conserved transcriptional responses.
For different species, the RNA-seq experiments will result in not only different read counts, but also different numbers and lengths of genes.  To  make the expression levels of orthologous genes comparable between different species, normalization is a crucial step  in the process of detecting  DE genes. This is in sharp contrast to the case of same species, where  the  numbers and lengths of genes are equal to each other. Therefore, the existing normalization
methods for same species may not provide a satisfactory performance or may not even be applicable for  RNA-seq data with different species. Therefore,
 developing new normalization methods for RNA-seq data with different species is extremely urgent.

In this paper, we propose a scale based normalization (SCBN) method between different species for RNA-seq data. Two main contributions of our work are:  (i)  dealing  with RNA-seq data with  two different species, which have different lengths of genes and sequencing depths, and (ii)  employing the hypothesis testing approaches to search  for the  optimal scaling factor, which minimizes the deviation between the empirical and   nominal type I errors.
From the simulation results, we find that the proposed SCBN method outperforms the existing median method, especially when the number of selected conserved genes is small or the selected conserved genes involves a lot of noise. In real data analysis, we analyze an  RNA-seq data of two species,  human and mouse, and  the results indicate that the SCBN  method delivers a more satisfactory performance than the median method.

Compared to the RNA-seq data with same species, normalization between different species is much more complicated. Although our proposed method has largely  improved the effectiveness to detect DE genes in some cases, we note that it may still not be able to provide a satisfactory performance when the rate of DE genes is very high in the whole samples. In addition,  the unmatched genes and the relation of orthologous genes  are not considered  in the process of normalization between different species.  This may call for a  future work  that develops   new methods to further improve our current method.

%


\begin{backmatter}

\section*{Funding}
Yan Zhou's research was supported by the National Natural Science Foundation of China (Grant No. 11701385), the National Statistical Research Project (Grant No. 2017LY56) and the Doctor Start Fund of Guangdong Province (Grant No. 2016A030310062).
Tiejun Tong's research was supported by the Hong Kong Baptist University grants FRG1/16-17/018 and FRG2/16-17/074, the Health and Medical Research Fund (Grant No. 04150476)  and the National Natural Science Foundation of China (Grant No. 11671338).
Bingqing Lin's research was supported by
the National Natural Science Foundation of China (Grant No. 11701386) and the Tianyuan Fund for Mathematics (Grant No. 11626159).

\section*{Availability of data and materials}
The data sets supporting the results of this article are included within the
article and the references. The ``SCBN" package is made in the form of R code and the complete documentation is available on
request from the corresponding author (zhangjunstat@gmail.com).
\section*{Author's contributions}
YZ, JDZ and JZ developed the SCBN method for normalization, conducted the simulation studies and real data analysis,
and wrote the draft of the manuscript.
JZ, TJT and BQL revised the manuscript.
JHW provided the guidance on methodology and finalized the manuscript.
All authors read and approved the final manuscript.
\section*{Competing interests}
The authors declare that they have no competing interests.
\section*{Consent for publication}
Not applicable.

\section*{Ethics approval and consent to participate}
Not applicable.


\bibliographystyle{bmc-mathphys} 

\section*{Additional Files}
  \subsection*{Additional file 1 --- The false discovery numbers at the rates of noise in selected conserved genes being 0, 0.1, 0.2, 0.3, 0.4 and 0.5, respectively.}
  \subsection*{Additional file 2 --- M versus A plots of two normalization methods.}
  \subsection*{Additional file 3 --- The scaling factors with different p-value cutoffs.}
  \subsection*{Additional file 4 --- Two groups of orthologous transcripts in human and mouse.}
  \subsection*{Additional file 5 --- The length difference of the orthologous transcripts between human and mouse.}
  \subsection*{Additional file 6 --- The histogram of the length difference of the orthologous transcripts between human and mouse.}
  \subsection*{Additional file 7 --- 143 most conserved orthologous transcripts  between human and mouse and orthology quality-controls criterion.}
  \subsection*{Additional file 8 --- $p$-values and $q$-values for each orthologous transcripts.}
  \subsection*{Additional file 9 --- The details for 140 and 123 differentially expressed orthologous transcripts detected by Median and SCBN method respectively.}

\end{backmatter}
\end{document}